\documentclass[aps,twocolumn,prl,showpacs]{revtex4}
\usepackage{times}
\usepackage{graphicx}
\usepackage{amsmath,amsfonts,bm,amssymb}
\usepackage{color}

\begin{document}

\title{Irreversible Monte Carlo Algorithms for Efficient Sampling}

\author{Konstantin S. Turitsyn$^{(a,b,c)}$, Michael Chertkov$^{(b,d)}$, Marija Vucelja$^{(d,b)}$}

\affiliation{$^a$ James Frank Institute, University of Chicago, Chicago, IL 60615, USA\\
$^b$ Center for Nonlinear Studies \& Theoretical Division, LANL, Los Alamos, NM 87545, USA\\
$^c$ Landau Institute for Theoretical Physics, Moscow 142432, Russia \\
$^d$ Department of Physics of Complex Systems, Weizmann Institute of Sciences, Rehovot 76100, Israel}

\begin{abstract}
Equilibrium systems evolve according to Detailed Balance (DB). This principe
guided development of the Monte-Carlo sampling techniques, of which
Metropolis-Hastings (MH) algorithm is the famous representative. It is also known
that DB is sufficient but not necessary. We construct irreversible deformation of a given reversible algorithm capable of dramatic improvement of sampling from known distribution.
Our transformation modifies transition rates keeping the structure of transitions intact. To illustrate the general scheme we design an Irreversible version of Metropolis-Hastings (IMH) and test it on example of a spin cluster. Standard MH for the model suffers from the critical slowdown, while IMH is free from critical slowdown.


\end{abstract}

\maketitle

Recent decades have been marked by fruitful interaction between physics and computer
science, with one of the most striking examples of such interaction going back to 40s when
physicists proposed a Markov Chain Monte Carlo (MCMC) algorithm \cite{MU49,MRRTT53}. MCMC
evaluates large sums, or integrals, approximately, in a sense imitating how nature would do efficient sampling itself. Development of this idea has became wide spread and proliferated a great variety of disciplines. (See \cite{LB00,Kra06,JS96} for a sample set of reviews in physics and computer science.) If one formally follows the letter of the original MCMC
suggestion one ought to ensure that the Detailed Balance (DB) condition is
satisfied. This condition reflects microscopic reversibility of the underlying equilibrium dynamics. A reader, impressed with indisputable success of the reversible MCMC techniques,
may still wonder if the equilibrium dynamics is the most efficient strategy for sampling and evaluating the integrals? In this letter we argue that typically the answer is NO.
Let us try to illustrate the ideas on a simple everyday life example. Consider mixing sugar in a cup of coffee, which is similar to sampling, as long as the sugar particles have to explore the entire interior of the cup. DB dynamics corresponds to diffusion taking an enormous mixing time. This is certainly not the best
way to mix. Moreover, our everyday experience suggests a better solution -- enhance mixing with a spoon. Spoon steering generates an out-of-equilibrium external flow which significantly accelerates mixing, while
achieving the same final result -- uniform distribution of sugar concentration over the cup.
In this letter we show constructively,  with a practical algorithm suggested,
that similar strategy can be used to decrease mixing time of any known reversible MCMC algorithms.

There are two main obstacles which prevent fast mixing by traditional MCMC methods.
First, the effective energy landscape can have high barriers, separating the energy
minima. In this case mixing time is dominated by rare processes of overcoming the
barriers. Second,  slow mixing can originate from the high entropy
of the states basin (too many comparably important states) providing major contribution to the system partition function. In the later case mixing time is determined by the number of steps it takes for reversible (diffusive) random walk to explore all the relevant states.

MCMC algorithms are best described on discrete example. Consider a graph with vertices
$i=1,\dots, {\cal N}$ each labeling a state of the system and edges $(i\to j)$
corresponding to  ``allowed'' transitions between the states. For instance,
an $N$-dimensional hypercube corresponds to a system of $N$ spins  (with ${\cal N} = 2^N$ states) with single-spin flips allowed.
An MCMC algorithm can be described
in terms of the transition matrix $T_{ij}$ representing the probability of a single MCMC step from
state $j$ to state $i$. Probability of finding
the system in state $i$ at time $t$, $P_i^t$, evolves according to the following Master Equation (ME):
$P_i^{t+1} = \sum_j T_{ij} P_j^t $. Stationary solution of ME, $P_i^t = \pi_i$, satisfies the Balance Condition (BC):
\begin{equation} \label{bal}
	\sum_j \left(T_{ij}\pi_j - T_{ji}\pi_i\right) = 0.
\end{equation}
$Q_{ij} = T_{ij}\pi_j$ from the lhs of Eq.~(\ref{bal}) can be interpreted as the stationary probability flux from state $i$ to state $j$. Obviously, stationarity of the probability flow reads as the condition for incoming and outgoing fluxes at any state to sum up to zero. Note also, that Eq.~(\ref{bal}) is nothing but incompressibility condition of the stationary probability flow.

The DB used in traditional MCMC algorithms is a more stringent condition, as it requires the piecewise balance of terms in the sum (\ref{bal}): for any pair of states with allowed transitions one requires, $T_{ij}\pi_j = T_{ji}\pi_i$. The main reason for DB to be so often used in practice originates from its tremendous simplicity. Otherwise, DB-consistent schemes constitute only a small subset of all other MCMC schemes convergent to the same stationary distribution $\pi$. From the hydrodynamic point of view reversible MCMC corresponds to irrotational probability flows, while irreversibility relates to nonzero rotational part, e.g. correspondent to vortices contained in the flow. Putting it formally, in the irreversible case antisymmetric part of the ergodic flow matrix is nonzero and it actually allows the following cycle decomposition, $Q_{ij} - Q_{ji} = \sum_\alpha J_\alpha(C_{ij}^\alpha - C_{ji}^\alpha)$,
where index $\alpha$ enumerates cycles on the graph of states with the adjacency matrices $C_{ij}^\alpha$. Then,  $J_\alpha$ stands for the magnitude of the probability flux flowing over cycle $\alpha$.

This cycle representation suggests how an irreversible MCMC algorithms can be constructed. One simply adds
cycles to a given reversible Markov Chain, ensuring that the transition probabilities remain
positive. In some simple examples this straightforward approach can be rather efficient.
Consider, for example, the problem of sampling uniform distribution on a two-dimensional lattice,
with the characteristic size  $L \gg 1$. Reversible Monte Carlo algorithms would require approximately
$T \sim L^2$ steps to converge. On the other hand, imposing a kind of super-lattice of size $l$ cycles, where $1 \ll l \ll L$,  one observes that typical mixing time is now determined by an
interplay of usual and ``turbulent'' diffusion respectively.  One estimates, $T \sim l^2 + L^2/l$, where the two terms correspond to dynamics on small (sub vortex-size) and large scales.
The minimal value of $T$ is achieved at $l \sim L^{2/3}$, thus resulting in $T \sim L^{4/3}$.
Moreover, one can reduce $T$ even further to $T\sim L$ with the help of an additional lifting operation, see e.g. \cite{CLP00} for related discussion. The important lesson we draw from this example is that knowing the state space and carefully planting irreversible cycles one can indeed achieve a significant acceleration of mixing.

However promising it looks,  the cycle-based procedure has two serious caveats making it difficult
to implement. First, the number of states in majority of interesting problems is (at least) exponential in the number of physical degrees of freedom. (For example an Ising system consisting of $N$ spins has $2^N$ states.) Thus, one expects that the number of cycles sufficient for essential mixing improvement is also exponential, i.e. too large of a number for algorithmically feasible implementation. Second, not all cycles are equally desirable, making optimization over cycle placements to be a task of even higher algorithmic complexity.

Therefore, aiming to achieve practical and flexible implementation, we have to abandon the cycle-based idea and instead focus on a better alternative -- building irreversible MCMC algorithms via controlled deformation of an existent reversible MCMC. To be more specific,  in the remainder of this Letter we adopt and develop replication/lifting trick discussed in \cite{DHN97,CLP00}.
The main idea behind our strategy is as follows. Instead of planting into the system an irreversible probability flux, correspondent to  an ``incompressible'' BC, we add a mixing desirable ``compressible'' flux, and compensate for its compressibility by building an additional replica with reversed flux and allowing some inter-replica transitions. To enforce BC one tunes the replica switching probabilities computed  ``on the fly'' (and locally). The replication idea is illustrated in Fig.~1. Acknowledging  generality of the setting, we focus in this Letter on explaining one relatively simple implementation of this idea.  Generalization and modifications of the procedure will be analyzed and discussed elsewhere.

\begin{figure}
	\centerline{
	\includegraphics[width=0.5\textwidth]{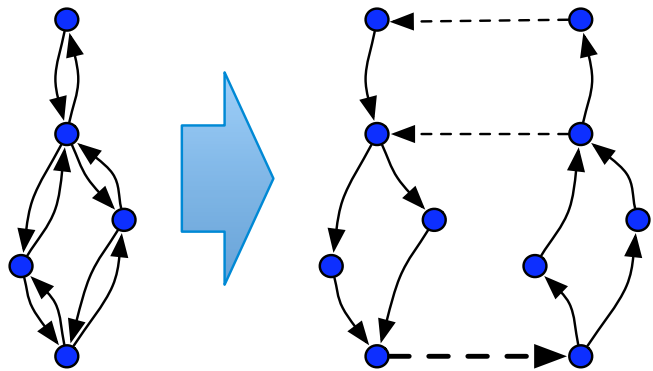} }
	\caption{Schematic representation of the replication deformation. Dashed lines represent replica switching transitions, which
compensate for compressibility of the probability flows associated with solid lines.
	} \label{figure:schematic}
\end{figure}

Consider reversible MCMC algorithm characterized by the transition matrix $T_{ij}$ which (a) obeys the DB condition, and (b) converges to the equilibrium distribution $\pi_i$. Assume that each state has duplicates in two replicas, marked by $\pm$. Following some local rule (an example will be provided below) one
introduces a split between states within each of the replicas, $T_{ij} = T_{ij}^{(+)} + T_{ij}^{(-)}$, such that all $T_{ij}^{(\pm)}$ are positive and satisfy, $\forall i\neq j$, $T_{ij}^{(+)} \pi_j = T_{ji}^{(-)} \pi_i$,  to be called the skew DB condition. The total transition matrix,
\begin{equation}
\hat{\cal T} = \left(\begin{array}{cccc}
\hat T^{(+)} & \hat \Lambda^{(+,-)} \\
\hat \Lambda^{(-,+)} &  \hat T^{(-)}
\end{array}\right),
\label{Tlift}
\end{equation}
also contains nonzero and positive (as probabilities) inter-replica terms, $\Lambda_{ii}^{(\pm,\mp)}$, allowing transitions only between two replicas of the same state.
One tunes the inter-replica terms to ensure convergence to the given steady distribution, $\pi_i$. This is achieved by choosing, $\Lambda_{ii}^{(\pm,\mp)} = \max\left\{0,\sum_j T_{ij}^{(\pm)} - T_{ij}^{(\mp)}\right\}>0$, and
the diagonal terms $T_{ii}^{(\pm)}$ are fixed according to the stochasticity condition:
$T_{ii}^{(\pm)} = 1 - \sum_j T_{ji}^{(\pm)} - \Lambda_{ii}^{(\mp,\pm)}$.
This description completes our construction of an irreversible MCMC algorithm from a given reversible
one. Note that this construction is not unique, and in general multiple choices of $\Lambda_{ii}^{(\pm,\mp)}$ are possible. The proposed scheme is illustrated below on example of a simple spin system, with the Metropolis-Hastings (MH)-Glauber algorithm chosen as the respective reversible prototype.

MH \cite{H70} is the most popular reversible MCMC algorithm.  MH-transition
from a current state $i$ is defined in two steps. (A) A new state $j$ is selected randomly.
(B) The proposed state is accepted with probability $p_{acc} = \min(1,\pi_j/\pi_i)$ or rejected with the
probability $1-p_{acc}$ respectively. Selecting the proposed state i.i.d. randomly from all possible single spin flips corresponds to the Glauber dynamics popular in simulations of spin systems. Let us now explain how to build an irreversible MCMC algorithm for spin systems based on the reversible MH-Glauber algorithm. One considers separation in two replicas according to the sign value, $+$ or $-$, of the spin to be flipped.  Then, our irreversible MH-Glauber scheme works as follows. Spin $\alpha$ is selected i.i.d. randomly from the pool of all other spins of the system having $+$ or $-$ values, depending on the sign of the replica where the system stays. The selected spin is flipped with the probability $p_{acc} = \min(1,\pi_j/\pi_i)$, in which case the system stays in the same replica. If the flip is not accepted the state is switched to its counterpart of the other replica with probability $\Lambda_{ii}^{(\mp,\pm)}/(1-\sum_j T_{ji}^{(\pm)})$. (These transitions are indicated as dash lines in Fig.~\ref{figure:schematic}.) Note, that in the case of the Glauber dynamics both $\Lambda_{ii}^{(\mp,\pm)}$ and $\sum_j T_{ji}^{(\pm)}$ are local
quantities depending only on the current state of the system, and calculating transition
probabilities constitutes an insignificant computational overhead.

We choose $N$-spins ferromagnetic cluster (equal strength interaction between all the spins) as a testbed and discuss sampling from respective stationary distribution, $\pi_{s_1...s_N} \sim\exp\left[(J/2N)\sum_{k,k'} s_k s_{k'}\right]$. Note,  that a state of the simple system is completely characterized by its global spin, $S=\sum_k s_k$, and respective probability distribution, $P(S)\sim\frac{N!}{N_{+}! N_{-}!} \exp\left[JS^2/(2N)\right]$, where
$N_\pm = (N\pm S)/2$ is the number of positive/negative spins. Considered in the thermodynamic limit, $N\to\infty$, the system undergoes a phase transition  at $J=1$. Away from the transition in the paramagnetic phase, $J<1$, $P(S)$ is centered around $S=0$ and the width of the distribution  is estimated by $\delta S\sim \sqrt{N/J}$, which changes to $\delta S\sim N^{3/4}$ at the critical point $J=1$. One important consequence of the distribution broadening is a slowdown observed at the critical point for reversible MH-Glauber sampling. Then characteristic correlation time of $S$ (measured in the number of Markov chain steps) is estimated as $T_{rev}\sim (\delta S)^2$, and the computational overhead associated with the critical slowdown is $\sim \sqrt{N}$. We brought this simple model to illustrate advantage of using irreversibility. As shown below, the irreversible modification of the MH-Glauber algorithm applied to the spin cluster problem achieves complete removal of the critical slowdown.
\begin{figure}
	\centerline{
	\includegraphics[width=0.5\textwidth]{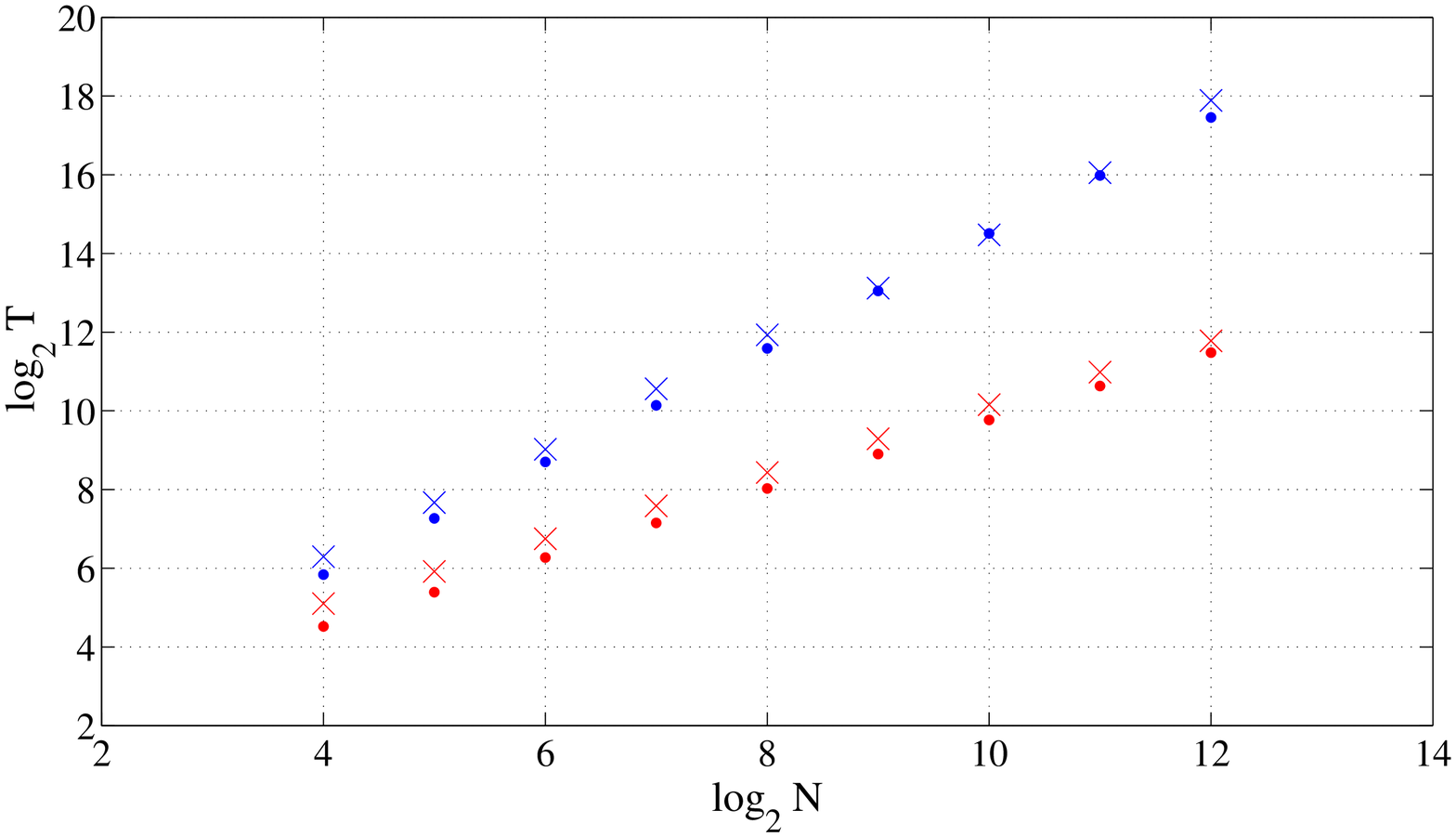} }
	\caption{Correlation time of the total spin de-correlation in the spin cluster model.
	Dots correspond to the direct diagonalization of the transition matrices.
Crosses are correlation times found from respective MCMC simulations. Blue and red colors correspond to reversible and irreversible algorithms respectively. Best fitting
	slopes are given by $T_{rev} \sim N^{1.43}$ and $T_{irr}\sim N^{0.85}$.
	} \label{figure:ctime}
\end{figure}
To estimate the correlation time in the irreversible case we first note that, switching from one replica to another the system always go through the $S=0$ state. ( This follows directly from the observation that $\Lambda_{ii}^{(+,-)} = 0$ for the states with $S>0$ and $\Lambda_{ii}^{(-,+)} = 0$ for the states with $S<0$.) The Markovian nature of the algorithm implies that all the trajectories connecting
two consequent $S=0$-swipes are statistically independent, and therefore correlation time is roughly
equal to the typical number of steps in each of these trajectories. Recalling that
inside a replica (i.e. in between two consecutive swipes) dynamics of $S$ is strictly monotonous, one estimates $T_{irr}\sim\delta S$. This estimate suggests a significant acceleration: $T_{irr} \sim \sqrt{T_{rev}} \ll T_{rev}$. Note, that one expects to observe significant acceleration even outside of the critical domain,  for larger and smaller values of $J$.

We verified the correlation time estimation via numerical tests. Implementing
reversible and irreversible versions of the MH-Glauber algorithm we, first, analyzed decay of
the pair correlation function, $\langle S(0) S(t)\rangle$, with time. Respective correlation time was reconstructed by fitting the large time asymptotics with exponential function, $\exp(-t/T_{rev})$, and exponential-oscillatory function, $\exp(-t/T_{irr})\cos(\omega t -\varphi)$, in the reversible and irreversible cases respectively. Second, for both MH and IMH algorithms we constructed 
transition matrix corresponding to the random walk in $S$, calculated spectral gap, $\Delta$, related to the correlation time as, $T = 1/{\mathrm Re \Delta}$. In both tests we analyzed critical point
$J=1$ and used different values of $N$ ranging from $16$ to $4096$. Simulation results are shown in Fig. \ref{figure:ctime}. The results found for two settings are consistent with each other. Numerical values ($T_{rev} \sim N^{1.43}$ and $T_{irr}\sim N^{0.85}$) are also in a reasonable agreement with respective theoretical predictions ($T_{rev} \sim N^{3/2}$ and $T_{irr} \sim N^{3/4}$) while a slight discrepancy can be attributed to finite size effects. Note, that in the irreversible case correlation time of the global spin correlation function (number of respective MC steps) grows with the number of spins, $N$, but does it slower than linearly. In other words, mixing becomes so efficient that equilibration of the global spin correlations is observed even before all spins of the systems are flipped. One concludes, that performance of the irreversible scheme is at least as good as the one of the cluster algorithms \cite{SW87,Wol89} tested on the
spin cluster model \cite{RTK89,PBKD96}. (We note, however, that direct comparison of the two algorithms is not straightforward, as the cluster algorithm flips many spins at once and therefore its convergence is normally stated in renormalized units.)

Let us now discuss relation of the proposed algorithm to previous studies. Although potential power of algorithms with broken DB has been realized for already a while, only handful of irreversible examples have been proposed so far. One of the examples is the sequential updating algorithm \cite{RO06} designed to simulate two-dimensional Ising system. In the essence, the algorithm consists of a number of subsystems (replicas) with internal dynamics, each characterized by its own transition matrix. In a great contrast with our algorithm, the system switches between replicas in a predefined deterministic fashion.  Similar idea of breaking DB by switching irreversibly but periodically between reversible portions was implemented in the successive over-relaxation algorithm  of \cite{A81}. Important sampling algorithm with DB broken is the Hybrid Monte Carlo of \cite{HK98}, where Hamiltonian dynamics is used to accelerate sampling. Once again the story here relates to replicas, each parameterized  by distinct momentum,  with switches between the replicas controlled deterministically by the underlying Hamiltonian. It is also appropriate to cite relevant efforts originated in statistics \cite{DHN97}, mathematics \cite{CLP00} and computer science \cite{JSS08}. Several simple examples of irreversible algorithms were discussed and analyzed in \cite{DHN97}. \cite{CLP00} showed that improvement in mixing, provided by a multi-replica lifting, does not allow reduction stronger than the one observed in the diffusive-to-ballistic scenario, $T \to \sqrt{T}$, where  $T$ is the mixing time of the underlying reversible algorithm. The grain of salt here is that the acceleration was achieved via a replication of an extremely high, $\sim k^2$, degree where $k$ is the number of states. \cite{JSS08} showed that some complementary ideas, from the field of distributed networks, allows to reduce this replication scaling a bit.

We also find it useful to briefly discuss reversible algorithms showing certain similarity to the algorithm and ideas of the Letter. First of all, it is important to mention again cluster algorithms \cite{SW87,ES88,Wol89} which were most successful in biting the odds of the critical slowdown in the regular systems of the Ising type. The trick here is to explore duality of the model, which allows two alternative representations related to each other via a state-non-local transformation. The cluster algorithm switches between two dual representations, thus realizing long jumps in the phase space. Note that such jumps would be forbidden by a phase-space local dynamics in either of the two representations. Best algorithms of the cluster type achieves very impressive rate of convergence. The downside is in the fact that the cluster algorithms are model specific and rather difficult in implementation because of extreme phase space non-locality of the steps. Worm algorithm of \cite{PS01} allows essential reduction in the critical slowdown via mapping to a high-temperature-inspired loop representation and making local moves there. The last but not the least, we mention the simulated annealing algorithm of \cite{KGV83} built on a temperature-graded replication consistent with DB. One interesting direction for future research is to explore if (and under which conditions) additional irreversibility can improve already good mixing performance provided within each of these reversible algorithms.

To summarize, this letter describes how to upgrade a reversible MC into an irreversible MC converging to the same distribution faster. To prove the concept we design a spin-problem specific irreversible algorithm,
and tested it on the mean-field spin-cluster model. We showed on this example that the irreversible modification can lead to dramatic acceleration of MC mixing. Our results suggest
that the irreversible MC algorithms are especially beneficial for
acceleration of mixing in systems containing multiple soft and zero modes, however unaccessible for standard (reversible) schemes. This situation occurs typically in systems experiencing
critical slowdown in the vicinity of a phase transition, and it is also an inherent property of systems
possessing internal symmetries of high degree. Entropic degeneracy is the main factor limiting the
convergence of regular MCMC algorithm in these problems. To conclude, the ideas discussed in the letter might be useful in studies of phase transitions, soft matter dynamics, protein structures and granular media.

The authors are grateful to V. Chernyak, F. Krzakala, J. Machta, D. Shah and T. Witten for inspiring discussions and useful remarks. The work at LANL was carried out under the auspices of
the National Nuclear Security Administration of the U.S. Department of Energy at Los
Alamos National Laboratory under Contract No. DE-AC52-06NA25396. MC also
acknowledges the Weston Visiting Professorship Program supporting his stay at the
Weizmann Institute, where part of this work was done.

\end{document}